\documentclass[%
 reprint,
 amsmath,amssymb,
 aps,
 prd,
 longbibliography
]{revtex4-1}
\usepackage{graphicx}
\usepackage{dcolumn}
\usepackage{bm}
\usepackage[hyperindex,breaklinks]{hyperref}
\usepackage{subfigure}
\usepackage{xcolor}
\hypersetup{
    colorlinks,
    linkcolor={red!50!black},
    citecolor={blue!50!blue},
    urlcolor={blue!80!black}
}

\begin{document}


\title{Simpler than vacuum: Antiscalar alternatives to black holes}

\author{Maxim A. Makukov}
\email{makukov@aphi.kz}
\author{Eduard G. Mychelkin}%
\email{mychelkin@aphi.kz}
\affiliation{%
 Fesenkov Astrophysical Institute \\
 050020, Almaty, Republic of Kazakhstan 
}%




\date{\today}

\begin{abstract}

The Janis-Newman-Winicour and Papapetrou metrics represent counterparts to the Schwarzschild black hole with scalar and antiscalar background fields, correspondingly (where ``anti'' is to be understood as in ``anti-de Sitter''). There is also a scalar counterpart (the Krori-Bhattacharjee metric) to the Kerr black hole. Here we study analytical connections between these solutions and obtain the exact rotational generalization of the antiscalar Papapetrou spacetime as a viable alternative to the Kerr black hole. The antiscalar metrics appear to be the simplest ones as they do not reveal event horizons and ergospheres, and they do not involve an extra parameter for scalar charge. Static antiscalar field is thermodynamically stable and self-consistent, but this is not the case for the scalar Janis-Newman-Winicour solution; besides, antiscalar thermodynamics is reducible to black-hole thermodynamics. Lensing, geodetic and Lense-Thirring effects are found to be practically indistinguishable between antiscalar and vacuum solutions in weak fields. Only strong-field observations might provide a test for the existence of antiscalar background. In particular, the antiscalar solution predicts 5\% larger shadows of supermassive compact objects, as compared to the vacuum solution. Another measurable aspect is the 6.92\% difference in the frequency of the innermost stable circular orbit,  characterizing the upper cut-off in the gravitational wave spectrum.

\end{abstract}

\maketitle


\section{Introduction}
\label{sec:introduction}

The field equations of General Relativity relate the energy-momentum tensor (EMT) ${T}_{\mu\nu}$ to the Einstein tensor ${G}_{\mu\nu}$ describing the geometry of spacetime, with the sign of the Einstein tensor to be chosen such that it conforms to observational data at Newtonian limit (the Poisson equation). However, for possible non-Newtonian background media with exotic equations of state the choice of the ${G}_{\mu\nu}$ sign should be made independently. E.g., the cosmological $\Lambda$-term, as a sort of background energy, for a given fixed value might manifest itself in two disguises~-- de Sitter and anti-de Sitter, both with the same equation of state $p = -\varepsilon$ implying that either energy density $\varepsilon$ or pressure $p$ is negative \cite{Weinberg1972}.

Similarly, minimal background scalar field $\phi$ with the equation of state $p=\varepsilon$ for timelike gradient $\partial_\mu \phi$ (or $p=-\varepsilon/3$ for spacelike $\partial_\mu \phi$) might also be related to positive or negative sign of the Einstein tensor, depending on the conformance to relevant experiments. We refer to these two alternatives as scalar and antiscalar cases. For the scalar case with spherically symmetric boundary conditions one obtains the solution known as the Janis-Newman-Winicour (JNW) metric \cite{Janis1968} (though it was found earlier, in a different form, by Fisher \cite{Fisher1948}). It reduces to the vacuum Schwarzschild metric in curvature coordinates when the scalar field vanishes; meanwhile, the corresponding rotational generalization of the JNW spacetime (the Krori-Bhattacharjee solution \cite{Krori1981}) reduces to the Kerr metric. For the antiscalar case one obtains the solution first found by Papapetrou \cite{Papapetrou1954b} and rediscovered later by Yilmaz \cite{Yilmaz1958}, and studied afterwards typically in the context of alternative theories of gravity (e.g.,  \cite{ROSEN1974455,itin1999gravity,robertson1999x,watt1999relativistic,zhuravlev1999latent,ibison2006cosmological,Aygun2008}); interestingly, recently it has been shown that this metric might be interpreted as a traversable wormhole \cite{boonserm2018exponential}. The rotational generalization of the exponential Papapetrou metric as antiscalar modification of the Kerr spacetime is obtained in this paper using two independent methods.

In case when the background is represented by the cosmological $\Lambda$-term, the standard way to study stability of the solution is by  perturbing the corresponding (de Sitter or anti-de Sitter) metric; the same goes for the vacuum Schwarzschild and Kerr solutions. In contrast, for a class of theories incorporating fundamental scalar background \cite{Mychelkin2015} the induced metric $g_{\mu\nu}=g_{\mu\nu}(\phi)$ loses an independent meaning as it is now determined by (anti)scalar field described by an additional (Klein-Gordon) equation. Since antiscalar metric coefficients happen to depend on $\phi$ smoothly (non-zero and well-behaved for all $r>0$), the problem of stability reduces, in effect,  to perturbing the scalar field in the corresponding Klein-Gordon equation.

One way to introduce perturbations is via incorporating a small mass-term into the Klein-Gordon equation considering the minimal field as a massless limit of some more realistic massive (anti)scalar field. In turn, for the latter, as discussed in Appendix~(\ref{sec:Stability}), there exist two possibilities in choosing the sign of the mass-term. As argued in \cite{Mychelkin2015}, for antiscalar field the negative mass-term is advisable, and in this case we find that the corresponding Klein-Gordon solution is stable, at least for large $r$. Furthermore, from the viewpoint of general-relativistic thermodynamics, the antiscalar field appears as thermodynamically stable and self-consistent medium, but this is not the case for the scalar JNW field -- see Appendix (\ref{sec:Thermod}), where it is also shown that anstiscalar thermodynamics contains black-hole thermodynamics as a particular case.

It stands to reason that changing the sign of the Einstein tensor as discussed above is formally equivalent to changing the sign of the corresponding EMT. Because the scalar EMT is quadratic in field, replacement of the scalar field by its antiscalar counterpart within such interpretation produces the following map: 
\begin{equation}
	{T}_{\mu\nu}^{\text{sc}}\left( \phi \right) \mapsto -{T}_{\mu\nu}^{\text{sc}}\left( \phi  \right) \quad \Leftrightarrow \quad  \phi \mapsto  i \phi,
	\label{ansatz}
\end{equation}
implying a similar map for the field source, i.e. the scalar charge $\sigma$: $\sigma \mapsto i \sigma$. This interpretation allows us to produce a new algorithm for the transformation of certain scalar-type metrics containing scalar charge into their antiscalar analogs and, thereby, to obtain new antiscalar solutions with subsequent application to observational effects, which will be covered in this paper. 

As a whole, we aim to study analytical and observational differences between stationary vacuum, scalar and antiscalar solutions, and pay special attention to the comparison of the newly obtained exact rotational generalization of the Papapetrou metric with the Kerr spacetime.

\section{Scalar-to-antiscalar transition} 
\label{sec:three_solutions}
We seek to compare three distinct physical situations~-- vacuum, scalar, and antiscalar, which are described by Einstein's equations with vacuum,  scalar and antiscalar minimal background, correspondingly:
\begin{eqnarray}
	\label{EE1}
	{G}_{\mu\nu} &=& 0, \\
	\label{EE2}
	{G}_{\mu\nu}& =& \varkappa {T}_{\mu\nu}^{\text{sc}}\left( \phi \right) ,\\
	\label{EE3}
	{G}_{\mu\nu} &= &-\varkappa {T}_{\mu\nu}^{\text{sc}}\left( \phi \right) ,
\end{eqnarray}
where $\varkappa=8 \pi G /c^4$ and the scalar field EMT is
\begin{equation}
	{T}^{\text{sc}}_{\mu\nu}(\phi) = \frac{1}{4 \pi} \left( \phi_\mu \phi_\nu - \tfrac{1}{2} {g}_{\mu\nu} \phi^\alpha \phi_\alpha \right), \,\,\,\, \phi_\mu \equiv \partial_\mu \phi.
	\label{EMT}
\end{equation}

In curvature coordinates, the spherically symmetric static solution of (\ref{EE1}) is the standard Schwarzschild metric (hereafter  we use units such that $G = c =1$):
\begin{equation}
	ds^2 =  \left(1-\frac{2M}{r}\right)dt^2 - \left(1-\frac{2M}{r}\right)^{-1} {dr}^2 -  r^2 d\Omega^2,
	\label{Schwarz}
\end{equation}
with $d\Omega^2 \equiv d\theta^2 + \sin^2 \theta d\phi^2$.

The JNW solution of (\ref{EE2}) might be represented in different forms \cite{[][ (in Russian; for English translation see \href{https://arxiv.org/abs/gr-qc/9911008}{arXiv:gr-qc/9911008})]Fisher1948,Janis1968,Agnese1985,Wyman1981,Virbhadra1997,bronnikov2011stability}; for our purposes we use the following one:
\begin{eqnarray}
	ds^2 = \left(1-\frac{2M}{\gamma  r}\right)^{\gamma }{dt}^2 &-&\left(1-\frac{2M}{\gamma  r}\right)^{-\gamma }{dr}^2 \nonumber\\
	&-& \left(1-\frac{2M}{\gamma  r}\right)^{1-\gamma } r^2 d\Omega^2,
	\label{JNW}
\end{eqnarray}
where $\gamma = M/\sqrt{M^2+ \sigma^2}$, and $\sigma$ is the scalar charge related to the solution of the corresponding (here, massless) Klein-Gordon equation:
\begin{eqnarray}
\Box \phi  & \equiv &  \frac{1}{\sqrt{-g}} \partial_\mu \left( \sqrt{-g} g^{\mu \nu} \partial_\nu \phi \right) =0 \quad \Rightarrow	\nonumber \\
 \phi &=& \frac{\sigma \gamma}{2M}\ln \left( 1-\frac{2M}{\gamma r}  \right)	\nonumber \\
 & =& \frac{1}{2} \sqrt{1- \gamma^2}\ln \left( 1 - \frac{2M}{\gamma r}  \right).
\label{KGsol}
\end{eqnarray}

The remarkable feature of the metric (\ref{JNW}) is that, due to the presence of the free parameter $\gamma$, it comprises solutions to both (\ref{EE1}) and (\ref{EE3}) as limiting cases. Thus, in the absence of scalar field $\sigma = 0$, hence $\gamma = 1$, and the JNW metric (\ref{JNW}) reduces to the Schwarzschild interval represented in curvature coordinates (\ref{Schwarz}), while (\ref{KGsol}) reduces identically to zero. 

On the other hand, following the ansatz (\ref{ansatz}), we obtain a physically reasonable result by applying in (\ref{JNW}) and (\ref{KGsol}) the transition $\sigma \mapsto  iM$, which is equivalent to 
\begin{equation}
\gamma=\frac{M}{\sqrt{M^2+ \sigma^2}} \to \infty.
\label{limit}
\end{equation} 
Application of this limit represents an algorithm which transforms (\ref{JNW}) directly into the antiscalar metric in isotropic coordinates:
\begin{equation}
	ds^2 =  e^{-2M/r} {dt}^2 - e^{2M/r} \left( {dr}^2 + r^2 d\Omega^2 \right),
\label{Papa}
\end{equation}
which satisfies Eq. (\ref{EE3}). This interval was first obtained by Papapetrou \cite{Papapetrou1954b}, who considered a class of metrics induced by scalar field, $g_{\mu\nu}=g_{\mu\nu}(\phi(x^{\alpha}))$, without reference to the EMT sign. Yilmaz later noted (in the footnote 4 of his paper \cite{Yilmaz1958}) that solution (\ref{Papa}) follows actually from the Einstein equations of the type (\ref{EE3}) rather than of the type  (\ref{EE2}).

Now, the corresponding limit (\ref{limit}) for the potential in (\ref{KGsol}) yields:
\begin{equation}
\lim_{\gamma \to \infty} \phi =   -i\frac{M}{r}, \quad \text{or} \quad \lim_{\gamma \to \infty} i\phi =   \frac{M}{r}.
\label{limphi}
\end{equation}
On the other hand, from (\ref{ansatz}) we have that $\phi \mapsto i \phi$, and so the antiscalar field proves to be nothing but the (positively defined) Newtonian potential
\begin{equation}
\phi = M/r,
\label{np}
\end{equation}
with the magnitude of scalar charge reduced to the central mass \footnote{If one rewrites equations (\ref{EE1})-(\ref{EE3}) as ${G}_{\mu\nu} = \epsilon \kappa {T}_{\mu\nu}$ with $\epsilon = \{+ 1,0, -1 \}$, then the general form of the solutions remains the same as in (\ref{JNW}) and (\ref{KGsol}), but now $\gamma = M/\sqrt{M^2 + \epsilon \sigma^2 }$. In this case for $\epsilon=-1$ and the condition of consistency with observations, $\sigma = M$, one again uniquely obtains (\ref{Papa}) and (\ref{np}).}.

The transfer from the JNW solution (\ref{JNW}) to the isotropic Papapetrou metric (\ref{Papa}) simultaneously transforms the system (\ref{EE2}) into (\ref{EE3}) and induces the transfer from scalar potential (\ref{KGsol}) to antiscalar potential (\ref{np}), and such operation is unique. This implies that all masses might be considered as sources of (anti)scalar field (cf. \cite{Bergmann1957}). Therefore, antiscalar field might represent a universal background unremovable from the Einstein equations, and whether it should be considered massless or massive  depends on the scales involved. E.g., massive antiscalar background expanded onto cosmological lengths with the mass-term estimated to be of order $m \approx 10^{-33}$ eV \cite{Mychelkin2015} might be potentially identifiable with dark energy phenomenon. At relatively small scales relevant to the topic of the present paper such mass-term is negligible.

\begin{figure}[t]
	\centerline{\includegraphics[width=8.6cm]{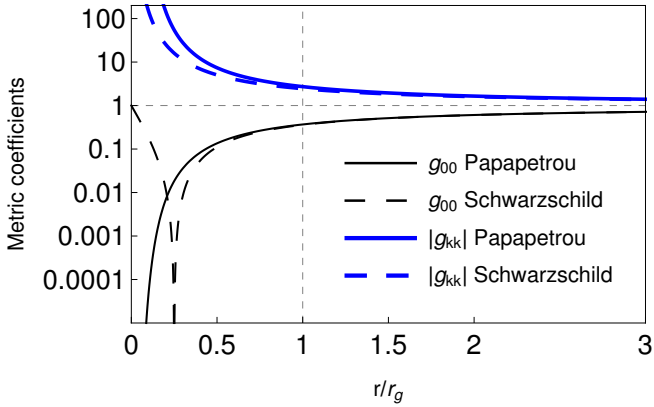}}
	\caption{Absolute values of the Papapetrou and Schwarzschild metric coefficients as functions of radial isotropic coordinate $r$ (normalized by $r_g=2M$).}
	\label{Fig:gcomp}
\end{figure}

Another peculiar feature of the metric (\ref{Papa}) is that it does not exhibit event horizon and thus does not represent a black hole. Nevertheless, the Papapetrou solution proves to be very similar in a number of respects to vacuum solution. The similarity between the two spacetimes is especially evident after recasting the Schwarzschild metric via the standard transformation 
\begin{equation}
r \mapsto r\left(1+\frac{M}{2r}\right)^2
\label{isotropTransform}
\end{equation}
from curvature coordinates to isotropic coordinates:
\begin{equation}
	ds^2 = \left(
	\frac{ 1-\frac{M}{2 r}}{1+\frac{M}{2 r}} \right)^2 {dt}^2 -
	\left(1+\frac{M}{2 r}\
		\right)^4
	\left( {dr}^2 +  r^2 d\Omega^2  \right).
	\label{SchwarzIsotropic}
\end{equation}
Then, the difference between (\ref{Papa}) and (\ref{SchwarzIsotropic}), even near the point $r_g=2M$, is practically negligible (see Fig.~(\ref{Fig:gcomp})). This guarantees conformity with the ``crucial'' effects of general relativity. 
We now turn to analyze other possible observational effects, including those arising in rotational generalizations of the vacuum and antiscalar solutions.

\section{Lensing effects}
\label{sub:Lens}
The role of scalar field in gravitational lensing was first considered by Virbhadra et al. \cite{Virbhadra1998}, with the key point that gravitational lensing might serve as a diagnostic tool for the scalar charge on the basis of the JNW solution, since it involves the integration constant naturally interpreted as scalar charge. In contrast, within antiscalar algorithm, the only justifiable non-zero choice for scalar charge magnitude is mass, as has been shown above.

\subsection{Light deflection} 
\label{sub:light_deflection}
First, we compare the light deflection angles for the antiscalar and vacuum cases, and demonstrate how both of those might be obtained from the JNW approach as corresponding limiting cases. The spherically symmetric interval might be written as
\begin{eqnarray}
ds^2 &= & g_{\alpha \beta}dx^\alpha dx^\beta = g_{tt}(r) {dt}^2 + g_{rr}(r) {dr}^2 \nonumber \\
& + & g_{\theta\theta}(r) {d\theta}^2 + g_{\phi\phi}(r,\theta) {d\phi }^2, 
\label{gensphmetric}
\end{eqnarray}
where $g_{\phi\phi}(r,\theta)=g_{\theta\theta}(r) \sin^{2}\theta$, and the signature ${(+---)}$ is absorbed into the metric components so that $g_{rr}$, $g_{\theta\theta}$ and $g_{\phi\phi}$ are  negative. For the closest distance of approach $r_0$, the exact deflection angle might be represented as follows \cite{Virbhadra1998,Virbhadra2002}:
\begin{widetext}
\begin{equation}
	\hat{\alpha}(r_0) = 2 \int_{r_0}^\infty{
	\left(\frac{g_{rr}(r)}{g_{\theta\theta}(r)} \right)^{1/2}
	\left( 
		\frac{g_{\theta\theta}(r)}{g_{\theta\theta}(r_0)} 
		\frac{g_{tt}(r_0)}{g_{tt}(r)} -1 
	 \right)^{-1/2}
	dr
	} - \pi,
\label{deflectionGeneral}
\end{equation}
yielding for the JNW metric (see \cite{Virbhadra1998,Virbhadra2002}):
\[
	\hat{\alpha}(r_0) = 2 \int_{r_0}^\infty{\frac{dr}
	{ r\sqrt{1- \frac{2M}{\gamma r}} 
	\sqrt{ \left( \frac{r}{r_0}  \right)^2 
	\left( 1 - \frac{2M}{\gamma r}  \right)^{1-2 \gamma}
	\left(  1 - \frac{2M}{\gamma r_0} \right)^{2 \gamma -1} -1
	  } 
	}} - \pi,
\]
\end{widetext}
or, up to the second order with respect to $M/r_0$,
\begin{eqnarray}
\hat{\alpha}(r_0) &= &  \frac{4M}{r_0} + \frac{4M^2}{r_0^2} \left( \frac{15 \pi}{16} -2  \right) \nonumber\\
 & + & \frac{2M^2}{r_0^2} \left(  \frac{2}{\gamma} - \frac{\pi (1- \gamma^2)}{8 \gamma^2}\right) + ...\,\,  .
\label{deflectionJNW1}
\end{eqnarray} 
The vacuum limit $\gamma=1$ gives (in  curvature coordinates)
\begin{equation}
\hat{\alpha}(r_0) = \frac{4M}{r_0} + \frac{4M^2}{r_0^2} \left( \frac{15 \pi}{16} - 1  \right) + ...\,\,  .
\label{deflectionSchwarz}
\end{equation}
This result might be also obtained directly by substituting (\ref{Schwarz}) into (\ref{deflectionGeneral}).

Now, following the same algorithm, we find from (\ref{deflectionGeneral}) and (\ref{Papa}) the deflection angle for the Papapetrou anti\-scalar metric (at the same order)
\begin{eqnarray}
	\hat{\alpha}(r_0) & = & 2 \int_{r_0}^\infty{\frac{dr}
	{ r
	\sqrt{ \left( \frac{r}{r_0}  \right)^2 
	e^{4M\left( \frac{1}{r} - \frac{1}{r_0} \right)} -1 } }} - \pi \nonumber\\
	&= & 
	\frac{4M}{r_0} + \frac{4M^2}{r_0^2} \left( \pi - 2  \right) + ...\,\,  ,
	\label{deflectionPapa}
\end{eqnarray}
where integration is performed with the method described in \cite{Weinberg1972}. Alternatively, this result might also be obtained as a limiting case $\gamma \to \infty$ from (\ref{deflectionJNW1}).

However, it should be stressed that direct comparison of results (\ref{deflectionSchwarz}) and (\ref{deflectionPapa}) would be incorrect, since the radial variables in these formulas are geometrically distinct as they label different types of coordinates (N.B.: for simplicity, we employ the same $r$-notation for radial distance in all coordinates, with concrete interpretation following from the context). To compare deflection angles (as well as other observational effects) for the Papapetrou and Schwarzschild spacetimes, one should use isotropic form of the Schwarzschild metric (\ref{SchwarzIsotropic}). Then, from (\ref{deflectionGeneral}) and (\ref{SchwarzIsotropic}) we get for vacuum solution:
\[
	\hat{\alpha}(r_0) = 2 \int\displaylimits_{r_0}^\infty{
	\frac{dr}{r}
	\left[
	\frac{\left( 2r + M  \right)^6}{ \left( 2r_0 +M  \right)^6 } 
	\frac{\left(2r_0 -M\right)^2}{ \left( 2r-M  \right)^2 }
	\frac{r_0^2}{r^2} - 1
	\right]^{-\frac{1}{2}} }
	 - \pi,
\]
and, following \cite{Weinberg1972}, after straightforward but cumbersome calculation up to the second order we obtain for isotropic vacuum case
\begin{equation}
\hat{\alpha}(r_0) = \frac{4M}{r_0} + \frac{4M^2}{r_0^2} \left( \frac{15 \pi}{16} - 2  \right) + ...\,\,  .
\label{deflSchwarzIsotrop}
\end{equation}
So, comparison of (\ref{deflectionPapa}) and (\ref{deflSchwarzIsotrop}) reveals somewhat more pronounced effect for the Papapetrou spacetime. However, the resulting difference is measurable in practice only in strong fields, and is negligible within the Solar system. To probe the difference, we need to turn to very massive objects like those found in the centers of galaxies, and to consider the imaging of their shadows.

\subsection{Shadows of compact objects} 
\label{sub:shadows_of_compact_objects}

For simplicity, we will restrict our consideration to the static case, leaving the effect of rotation on the shadow image for later studies. As a first step, we calculate the impact parameter of the so-called photon sphere. The general impact parameter $J= D_l \, \text{sin}\Theta $ (here $D_l$ is distance from the observer to the lens and $\Theta$ is the observer's polar angle for image of the source) represented in terms of the general metric (\ref{gensphmetric}),
\begin{equation}
	J = J(r_0) =\left( \frac{-g_{\theta\theta}(r_0)}{g_{tt}(r_0)} \right)^{1/2},
	\label{J}
\end{equation}
for the JNW-metric becomes \cite{Virbhadra1998,Virbhadra2002}
\begin{equation}
	J= r_0 \left(1-\frac{2M}{r_0 \gamma} \right)^{(1-2\gamma)/2}.
	\label{JJNW}	
\end{equation}
From here, for vacuum ($\gamma=1$) in  curvature coordinates:
\begin{equation}
	J= r_0 \left(1-\frac{2M}{r_0 } \right)^{-1/2} = 1 + \frac{M}{r_0} + O\left(\frac{M^2}{r_0^2}\right),
	\label{JSchwarzCurv}
\end{equation}
while for antiscalar background ($\gamma \to \infty$) we get
\begin{equation}
	J= r_0 \exp {\left(\frac{2M}{r_0 } \right)} = 1 + \frac{2M}{r_0} + O\left(\frac{M^2}{r_0^2}\right).
	\label{JPapa}
\end{equation}
As for the Schwarzschild isotropic coordinates, from (\ref{J}) it follows:
\begin{equation}
	J=\frac{r_0 \left(1+\frac{M}{2 r_0}\right)^3}{1-\frac{M}{2 r_0}} = 1 + \frac{2M}{r_0} + O\left(\frac{M^2}{r_0^2}\right).
	\label{JSchwarzIsotrop}
\end{equation}
At first order, expressions (\ref{JSchwarzIsotrop}) and (\ref{JPapa}) coincide but differ from (\ref{JSchwarzCurv}). 

The characteristic related to the shadow imaging -- the photon sphere -- arises when the deflection angle (\ref{deflectionGeneral}) is maximized, which means that the derivative of (\ref{deflectionGeneral}) with respect to $r_0$ is zero, i.e.
\begin{equation}
	g_{tt}\frac{\partial g_{\theta \theta}}{\partial{r_0}}=g_{\theta \theta}\frac{\partial g_{tt}}{\partial{r_0}} \, .
	\label{phsph}
\end{equation}
The solution of this equation yields the radius of photon sphere $r_0=r_\text{ps}$. Then for scalar JNW-metric (\ref{JNW}) one obtains:
\begin{equation}
r_\text{ps}=\frac{M}{\gamma}(1+2\gamma).
\label{rps}
\end{equation}
So, in vacuum with curvature coordinates $r_\text{ps}=3M$, and for antiscalar case we get $r_\text{ps}=2M$, while for isotropic Schwarzschild's coordinates we obtain four solutions of (\ref{phsph}):
$$r_\text{ps}=	\pm \frac{M}{2}, \qquad  r_\text{ps}=\left(1\pm\frac{\sqrt{3}}{2}\right)M,$$
and the only solution which does not lead to contradictions is the maximal positive one:
\begin{equation}
	r_\text{ps}=\left(1+\frac{\sqrt{3}}{2}\right)M \approx 1.866M.
	\label{phsphIsotrop}
\end{equation}
After substitution of the JNW photon sphere radius (\ref{rps}) instead of $r_0$ in (\ref{JJNW}), the resulting impact parameter (which characterizes the radius of the shadow: $J(r_\text{ps})=R_\text{sh}$, see, e.g., \cite{Tsupko2018}) proves to be \cite{Virbhadra1998,Virbhadra2002}:
\begin{equation}
J(r_\text{ps})= R_\text{sh} = M \frac{1+2 \gamma}{\gamma} \left( 1 - \frac{2}{1+2\gamma}  \right)^{\frac{1}{2}-\gamma} \, ,
\label{JpsJNW}
\end{equation}
i.e. for vacuum in curvature coordinates ($\gamma=1$) it follows that $R_\text{sh}= 3\sqrt {3} M = 5.196 M$.  Remarkably, in the Schwarzschild isotropic coordinates, substituting (\ref{phsphIsotrop}) into (\ref{JSchwarzIsotrop}), we get exactly the same value for the shadow size, $R_\text{sh}= 3 \sqrt{3} M.$ This may be comprehensible because geometrically different photon spheres represent SO(3)$\times\mathbb{R}$-invariant surfaces \cite{Claudel2001} describing propagation of photons around black holes. 

At the same time, for the antiscalar case ($\gamma \to \infty$) from (\ref{JpsJNW}) we find $R_\text{sh}= 2eM=5.437 M$, which corresponds to a physically distinct situation and proves to be 5\% larger than in the vacuum case. 

Thus, we have two different expectations for the shadow size for the same value of the central mass $M$. The mass of compact objects is measurable independently, e.g., via surrounding orbits of test particles, and so the observed size of the shadow might distinguish between the vacuum and antiscalar cases.

Next, we turn to spin precession effects in vacuum and antiscalar backgrounds.

\section{General spin precession}
\label{GenPrecEff}
The general  frequency $\Omega$ of a test gyro in an arbitrary stationary spacetime with a timelike Killing vector $K$ can be expressed in terms of differential forms \cite{Chakraborty2017a} as
\begin{equation}
	\tilde{\Omega} = \frac{1}{2 K^2}\ast \left(\tilde{K} \wedge d\tilde{K}  \right),
\label{oneforms}
\end{equation}
where $\tilde{\Omega}$ and $\tilde{K}$ are the one-forms of $\Omega$ and $K$, and $\ast$ represents Hodge dual. The $K$ vector might be represented as a linear combination of time-translational and azimuthal vectors, $K = \partial_t +\omega\partial_\phi,$ where $\omega$ is the angular velocity for an observer moving along integral curves of the $K$-field \cite{Chakraborty2017a}. With the coordinate-free form of spin precession (\ref{oneforms}), the vector field corresponding to general precession rate may be represented as (see \cite{Chakraborty2017a}):
\begin{widetext}
\begin{eqnarray}
 \vec{\Omega}&=& \sqrt{-g_{rr}} \Omega^r \hat{r} + \sqrt{-g_{\theta\theta}} \Omega^\theta \hat{\theta} = \frac{1}{2\sqrt {-g}\left(1+2\omega\frac{g_{t\phi}}{g_{tt}}+\omega^2\frac{g_{\phi\phi}}{g_{tt}}\right)} \nonumber \\
 &\times& \left\{
 \sqrt{-g_{rr}}\left[\left(g_{t\phi,\theta}
-\frac{g_{t\phi}}{g_{tt}} g_{tt,\theta}\right)+\omega\left(g_{\phi\phi,\theta}
-\frac{g_{\phi\phi}}{g_{tt}} g_{tt,\theta}\right)+ \omega^2 \left(\frac{g_{t\phi}}{g_{tt}}g_{\phi\phi,\theta}
-\frac{g_{\phi\phi}}{g_{tt}} g_{t\phi,\theta}\right) \right]\hat{r} \right. \nonumber\\
 & &- \left. \sqrt{-g_{\theta\theta}} \left[ \left( g_{t\phi,r} -\frac{g_{t\phi}}{g_{tt}} g_{tt,r} \right) + \omega \left( g_{\phi\phi,r} -\frac{g_{\phi\phi}}{g_{tt}} g_{tt,r} \right) + \omega^2 \left( \frac{g_{t\phi}}{g_{tt}}g_{\phi\phi,r} -\frac{g_{\phi\phi}}{g_{tt}} g_{t\phi,r} \right) \right] \hat{\theta} \right\}.
\label{GenPrec}
\end{eqnarray}
\end{widetext}
The magnitude of this vector is
\begin{equation}
	\Omega(r,\theta) = |\vec{\Omega}| = \sqrt{  -g_{rr} (\Omega^r)^2 - g_{\theta\theta} (\Omega^\theta)^2 }
	\label{magnitude}
\end{equation}
and will be used in subsequent calculations.

\section{Geodetic precession} 
\label{sec:geodetic}

\subsection{JNW (scalar) case} 
\label{sub:jnw_geodetic}

 Substitution of the JNW metric (\ref{JNW}) into  (\ref{GenPrec})-(\ref{magnitude}) leads to the general precession frequency for gyro moving  in equatorial plane ($\theta=\pi/2$) with orbital angular speed $\omega$:

\begin{equation}
	\Omega =\frac{\omega   \left(1 -\frac{2 M}{r}-\frac{M}{\gamma  r}\right)\left(1-\frac{2 M}{\gamma  r}\right)^{-(\gamma+3)/2}}{\left(1-\frac{2 M}{\gamma  r}\right)^{-1}-r^2 \omega ^2 \left(1-\frac{2 M}{\gamma r} \right)^{-2 \gamma}}.
	\label{OmegaJNW}
\end{equation}

Since it appears that analysis of the geodetic effect for the JNW metric is absent in the literature, here we present its sufficiently full derivation. It is common to choose circular geodesics with corresponding angular velocities, for which we  denote $\omega = \omega_c$ and $r=R=const$. Traditionally, the circular frequencies $\omega_c$ might be obtained from the standard Hamilton-Jacobi formalism \cite{Padmanabhan2010}. Following that procedure, we start from the Hamilton-Jacobi equation for a test body of mass $m$,
\begin{equation}
	g^{\mu \nu} \frac{\partial S}{\partial x^\mu} \frac{\partial S}{\partial x^\nu} - m^2  =0,
	\label{HJ}
\end{equation}
written for the JNW metric:
\begin{equation}
	\left( \frac{\partial S}{\partial t}  \right)^2 B^{-\gamma}
	 - \left( \frac{\partial S}{\partial r}  \right)^2 B^\gamma
	 - \left( \frac{\partial S}{\partial \phi}  \right)^2 \frac{B^{\gamma-1}}{r^2} 
	  = m^2,
	\label{HJ-JNW}
\end{equation}
with $B =1 - 2M /(\gamma r)$. Due to spherical symmetry, we look for the action in the form (see, e.g., \cite{Padmanabhan2010})
\[
	S = -E t + L \phi + S_r(r),
\]
where $E$ and $L$ are constant energy and angular momentum, respectively, and $S_r$ is the part of the action that depends only on $r$. The solution to (\ref{HJ-JNW}) is then
\[
	S = - E t + L \phi +
	\int{ \sqrt{E^2 B^{-2 \gamma} - {L^2}{r^{-2}} B^{-1} - m^2 B^{-\gamma}} dr }.
\]
Taking derivatives of this with respect to $E$ and $L$ and equating them to constants, we find
\[
	\frac{\partial\phi}{\partial r}  = \frac{L}{B r^2}\left(  E^2 B^{-2 \gamma} - {L^2}{r^{-2}} B^{-1} - m^2 B^{-\gamma} \right)^{-{1}/{2}},
\]
\[
	\frac{\partial r}{\partial t} = \frac{B^{2 \gamma}}{E} \left(  E^2 B^{-2 \gamma} - {L^2}{r^{-2}} B^{-1} - m^2 B^{-\gamma} \right)^{1/2},
\]
so
 the orbital angular velocity   $\omega = \partial \phi / \partial t$ is:
\[
	\omega = \frac{L B^{2 \gamma -1}}{E r^2}. 
\]
For circular orbits, expressing $L/E$  in a standard way (based on the transformation to a new  variable $u = 1/r$), it follows that
\[
	\omega_c = \left( \frac{M}{R^3} \right)^{\frac{1}{2}}  \left( 1 - \frac{2M}{\gamma R} \right)^{\gamma - \frac{1}{2}} \left(  1- \frac{M}{R} - \frac{M}{\gamma R} \right)^{-\frac{1}{2}} .
\]
Substituting this result into (\ref{OmegaJNW}), we obtain the corresponding gyro precession frequency (with respect to proper time):
\[
	\Omega= \left( \frac{M}{R^3} \right)^{\frac{1}{2}} \left( 1- \frac{2M}{\gamma R} \right)^{\frac{\gamma-2}{2}} \left( 1- \frac{M}{R} - \frac{M}{\gamma R} \right)^{\frac{1}{2}}. 
\]
To express this quantity in terms of coordinate time, we need the relation between coordinate and proper times, $u^t = dt/d\tau$. Then, from $u^i u_i=1$, with $u^\phi = \omega_c u^t$, $u^r = u^\theta = 0$, we find for the JNW metric:
\begin{equation}
	u^t =  \left( 1- \frac{2M}{\gamma R} \right)^{-\frac{\gamma}{2}} \left(  1 - \frac{M}{R \left(  1- \frac{M}{R} - \frac{M}{\gamma R} \right) }  \right)^{-\frac{1}{2}},
	\label{utII}
\end{equation}
so the gyro precession frequency with respect to coordinate time is:
\begin{equation}
	\Omega' =  \frac{\Omega}{u^t} =  \omega_c \sqrt{
		\frac{\left( 1-\frac{M}{\gamma R} -\frac{M}{R} \right) \left( 1 - \frac{2M}{R} - \frac{M}{\gamma R}  \right) }
		{1 - \frac{2M}{\gamma R}} }.
	\label{precfrII}
\end{equation}
$\Omega'$ is always directed opposite to $\omega_c$ \footnote{This statement is valid for circular orbits in any static spherically symmetric metric, as follows from the solution of the general spin-transfer equation \cite{Padmanabhan2010}.}, and the resulting difference between their magnitudes produces the geodetic precession effect. Thus, during the time interval $t = 2 \pi / \omega_c$ the direction of spin changes by the angle
\begin{equation}
	\alpha	= \frac{2 \pi}{\omega_c} \left(\omega_c - \Omega' \right) ,
	\label{accumAngle}
\end{equation}
and for (\ref{precfrII}) this yields:
\begin{equation}
	\alpha =  2 \pi  \left(1-  \sqrt{
			\frac{\left( 1-\frac{M}{\gamma R} -\frac{M}{R} \right) \left( 1 - \frac{2M}{R} - \frac{M}{\gamma R}  \right) }
			{1 - \frac{2M}{\gamma R}} }\right).
	\label{alfaJNW}
\end{equation}
Adopting $\gamma = 1$ we obtain the corresponding results for vacuum solution \emph{in curvature coordinates} (these results are known and might be found elsewhere, see, e.g., \cite{Padmanabhan2010,Chakraborty2017a}). Another limit we are interested in, $\gamma \to \infty$, yields antiscalar results \emph{in isotropic coordinates}. For further comparison, we need to derive corresponding results for the vacuum solution in isotropic coordinates as well. These might be obtained either by  applying the transformation (\ref{isotropTransform}) to the mentioned results \cite{Padmanabhan2010,Chakraborty2017a}, or starting from scratch. For completeness, we provide brief direct derivation.

\subsection{Schwarzschild (vacuum) case in isotropic coordinates} 
\label{sub:SchwarzIs_metric}

Substitution of the Schwarzschild isotropic metric (\ref{SchwarzIsotropic}) into  (\ref{GenPrec})-(\ref{magnitude}) leads to the precession frequency for gyro moving  in equatorial plane with $\theta=\pi/2$ with some orbital angular velocity $\omega$:
\begin{equation}
	\Omega =\frac{\omega \left(1-\frac{2 M}{r}+\frac{M^2}{4 r^2}\right)}{\left(1-\frac{M}{2 r}\right)^2-r^2 \omega^2 \left(1+\frac{M}{2 r}\right)^6} .
	\label{OmegaSchwarzIs}
\end{equation}
For circular geodesics with $r=R=const$ and corresponding orbital angular velocity $\omega = \omega_c$ we obtain, following the standard procedure of solving the Hamilton-Jacobi equation: 
\begin{equation}
	\omega_c = \sqrt{\frac{M}{R^3}} \left(1+\frac{M}{2 R}\right)^{-3}.
	\label{w0SchwarzIs}
\end{equation}
Now, substituting (\ref{w0SchwarzIs}) into (\ref{OmegaSchwarzIs}) one obtains the gyro precession frequency (with respect to proper time) the magnitude of which exactly coincides with that of (\ref{w0SchwarzIs}):
\[
	\Omega = \sqrt{\frac{M}{R^3}} \left(1+\frac{M}{2 R}\right)^{-3} . 
\]
The relation between coordinate and proper times $u^t = dt/d\tau$ for a gyro moving on a circular orbit in this spacetime is found as
\[
	u^t = \left(1+\frac{M}{2 R}\right) \left( 1 -\frac{2 M}{R} + \frac{M^2}{4 R^2} \right)^{-1/2}.
\]
So, the final gyro precession frequency with respect to coordinate time is:
\begin{equation}
	\Omega'= \frac{\Omega}{u^t} = \omega _c  \left(1 + \frac{M}{2 R}  \right)^{-1} \left( 1-\frac{2 M}{R}+\frac{M^2}{4 R^2}  \right)^{1/2},
	\label{precfrIIs}
\end{equation}
and from (\ref{accumAngle}) one obtains per one orbital revolution:
\begin{equation}
	\alpha = 2 \pi \left[ 1 -  \left(1 + \frac{M}{2 R}  \right)^{-1} \left( 1-\frac{2 M}{R}+\frac{M^2}{4 R^2}  \right)^{1/2} \right].
	\label{anglprecIIs}
\end{equation}
Evidently, this formula is applicable for distances above the point $R = M\left(1+\sqrt{3}/2\right)$, where $\Omega$ in (\ref{OmegaSchwarzIs}) changes sign.

\subsection{Papapetrou (antiscalar) case} 
\label{sub:papapetrou_}
Substitution of the metric (\ref{Papa}) into  (\ref{GenPrec})-(\ref{magnitude}) leads to precession frequency for gyro moving  in equatorial plane with orbital angular velocity $\omega$:
\begin{equation}
	\Omega =\frac{\omega  \left(1-2 M/r\right) e^{M/r}}{1-r^2 \omega ^2 e^{\frac{4 M}{r}}}
	\label{OmegaPapa}
\end{equation}
(as might be checked, this expression also follows as a limit from the scalar counterpart (\ref{OmegaJNW}) with $\gamma \rightarrow \infty$).

To consider the geodetic effect for antiscalar case, we again choose circular geodesics and follow the Hamilton-Jacobi formalism as in the previous section. In this case, the action equation (\ref{HJ}) becomes:
\[
	\left( \frac{\partial S}{\partial t}  \right)^2 e^{2M/r} - \left( \frac{\partial S}{\partial r}  \right)^2 e^{-2M/r} - \left( \frac{\partial S}{\partial \phi}  \right)^2 \frac{e^{-2M/r}}{r^2} = m^2.
\]
Its general solution is 
\[
	S = - E t + L \phi + \int{ \sqrt{E^2 e^{4M/r} - {L^2}{r^{-2}} - m^2 e^{2M/r}} dr }.
\]
Taking derivative with respect to $E$ and $L$ and equating them to constants, we find $\partial\phi/\partial r$ and $\partial r/ \partial t$ and, ultimately, the orbital angular velocity $\omega = \partial \phi / \partial t$:
\[
	\omega = \frac{L e^{-4M/r}}{E r^2}. 
\]
For circular orbits with radius $R$, expressing $L/E$  in the standard way \cite{Padmanabhan2010}, we find
\begin{equation}
	\omega_c = \sqrt{\frac{M}{R^3} }  \left( 1- \frac{M}{R}  \right)^{-{1}/{2}} e^{-2M/R}.	
	\label{wcirc}
\end{equation}
Substituting this into (\ref{OmegaPapa}), one obtains the gyro precession frequency (with respect to proper time):
\[
	\Omega = \sqrt{ \frac{M}{R^3} } \left( 1- \frac{M}{R} \right)^{1/2} e^{-{M}/{R}}.
\]
Next, instead of (\ref{utII}) it may be found that
\[
	u^t =  \frac{\partial t}{\partial \tau} =  \left(  \frac{ 1-M/R  }{1-2M/R}   \right)^{1/2}   e^{M/R},
\]
and so the gyro precession frequency with respect to coordinate time is:
\begin{equation}
	\Omega' = \frac{\Omega}{u^t} = \omega_c \sqrt{1- \frac{3M}{R}+ \frac{2M^2}{R^2}}.
	\label{precfrIII}
\end{equation}

Now, from (\ref{accumAngle}) and (\ref{precfrIII}), the angle of precession during one revolution of gyro on a circular orbit becomes:
\begin{equation}
	\alpha = 2 \pi \left( 1 - \sqrt{1- \frac{3M}{R}+ \frac{2M^2}{R^2}}    \right),
	\label{anglprecIII}
\end{equation}
which also follows from the limit $\gamma \to \infty$ in (\ref{alfaJNW}).

Similar to the Schwarzschild case, the result in (\ref{anglprecIII}) is restricted to $R \ge 2M$, and the expression for $\Omega$ in (\ref{OmegaPapa}) changes sign for $R=2M$.
\begin{figure}
	\centerline{\includegraphics[width=8.6cm]{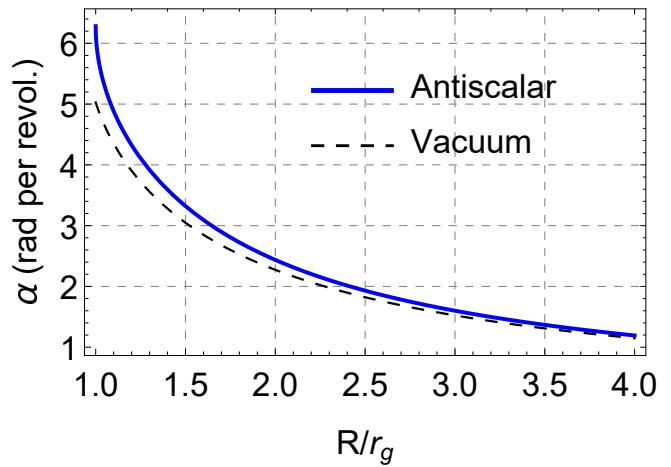}}
	\caption{Geodetic precession angle $\alpha$ per one orbital revolution as a function of circular orbit radius $R$ for vacuum and antiscalar solutions, both in isotropic coordinates.}
	\label{Fig:geodetic}
\end{figure}

Accumulated during large number of periods, both geodetic effects, antiscalar  (\ref{anglprecIII}) and vacuum (\ref{anglprecIIs}), might be measurable in satellite experiments \cite{Everitt2011}. However, for realistic measurement times, the difference between those might become significant only in strong field regime at distances comparable to $r_g$, as can been seen in Fig. (\ref{Fig:geodetic}).

\subsection{Upper limit in the gravitational wave spectrum} 
\label{sub:gw}
The frequency of circular orbits (\ref{wcirc}) is also relevant in the study of gravitational radiation emitted during inspiral mergers. As shown by Watt \& Misner \cite{watt1999relativistic} the effective potential of the Papapetrou metric (\ref{Papa}) has a ``pit'', just like in the Schwarzschild case, which leads to the existence of the innermost stable circular orbit (ISCO) with radius $R_\text{isco} = M(3+\sqrt{5})$. Substituted into (\ref{wcirc}), this yields the frequency of ISCO,
\begin{equation}
\omega_\text{isco} = \frac{0.0633326}{M},
\label{isco}	
\end{equation}
which determines the upper limit in the observed gravitational wave spectrum. Watt \& Misner, however, compare this with the corresponding known vacuum analogs in \emph{curvature} coordinates ($R_\text{isco}^\text{curv}=6M$, $\omega_\text{isco}^\text{curv} = 0.0680414/M$), yielding the $6.92\%$ difference as compared to antiscalar case (\ref{isco}).

To perform double-check we use isotropic coordinates \emph{ab initio} in both antiscalar and vacuum cases. Then, for the Schwarzschild metric in isotropic coordinates we obtain $R_\text{isco}^\text{isot}=\left(\frac{5}{2}+\sqrt{6}\right)M \approx 4.9495M$ which, after substitution into (\ref{w0SchwarzIs}), produces exactly the same numerical value for $\omega_\text{isco}^\text{isot} = 0.0680414/M$. In fact, this is not surprising, since the transformation (\ref{isotropTransform}) from curvature to isotropic form does not involve angular and time coordinates. So, indeed, the frequencies of innermost stable circular orbits characterizing cut-off in gravitational wave spectrum differ by 6.92\% between vacuum and antiscalar cases.

As a next step, we compare effects of the central mass rotation in vacuum, scalar and antiscalar field backgrounds.

\section{Lense-Thirring effect}

For $\omega=0$, expression (\ref{GenPrec}) reduces to the Lense-Thirring precession \cite{Chakraborty2017a,0264-9381-31-7-075006}:
\begin{eqnarray}
 \vec{\Omega}|_{\omega=0} &=& \vec{\Omega}_\text{LT} = \sqrt{-g_{rr}}\Omega_\text{LT}^r \hat{r} + \sqrt{-g_{\theta\theta}} \Omega_\text{LT}^\theta \hat{\theta}=  \nonumber \\
  &=&  \frac{1}{2\sqrt {-g}}\left[\sqrt{-g_{rr}}\left(g_{t\phi,\theta}
-\frac{g_{t\phi}}{g_{tt}} g_{tt,\theta}\right)\hat{r} \right. \nonumber\\
& &-\left. \sqrt{-g_{\theta\theta}}\left(g_{t\phi,r}-\frac{g_{t\phi}}{g_{tt}}
g_{tt,r}\right)\hat{\theta}\right].
\label{GenPrec0}
\end{eqnarray}
The magnitude of this vector,
\begin{equation}
	\Omega_\text{LT}(r,\theta) = |\vec{\Omega}_\text{LT}| = \sqrt{  -g_{rr} (\Omega_\text{LT}^r)^2 - g_{\theta\theta} (\Omega_\text{LT}^\theta)^2 },
	\label{magnitudeLT}
\end{equation}
will be used in subsequent calculations.

\subsection{Rotating mass in the scalar background} 
\label{sec:rotJNW}
The rotational generalization of the JNW metric was first obtained in \cite{Krori1981},  and, later and in a different form, rediscovered in \cite{Agnese1985}. We will use the solution in the simple form as presented, e.g., in \cite{Gyulchev2008}:
\begin{eqnarray}
ds^2 &=& \left(1 - \frac{A}{\gamma}\right)^\gamma \left(dt-W d\phi \right)^2 \nonumber\\ 
&-& \left(1 - \frac{A}{\gamma}\right)^{1-\gamma} \rho^2\left(\frac{dr^2}{\Delta}+d\theta^2+\sin^2\theta d\phi^2\right) \nonumber\\
&+& 2W(dt-W d\phi)d\phi,
\label{RJNW}
\end{eqnarray}
where 
\begin{eqnarray}
A=\frac{2M r}{\rho^2}, \quad  \rho^2 = r^2 +a^2 \cos^2\theta, \nonumber\\
 W=a\sin^2\theta,  \quad \Delta=r^2 + a^2 - \tfrac{2M r}{\gamma},
\label{defs}
\end{eqnarray}
and $a$ is the specific rotation parameter.
The Jacobian for the metric (\ref{RJNW}) is
\[
	\sqrt{-g}= \rho^2  \left(1 - \frac{A}{ \gamma}  \right)^{1- \gamma}  \sin\theta.
\]

The corresponding solution of the Klein-Gordon equation can be found as
\begin{equation}
	\phi = \frac{1}{2}\sqrt{1- \gamma^2}\ln\left(1-\frac{A}{\gamma}\right),
	\label{KGRJNW}
\end{equation}
which for $a=0$ reduces to (\ref{KGsol}).

In accord with (\ref{magnitudeLT}), vector (\ref{GenPrec0}) evaluated for the metric (\ref{RJNW}) has the magnitude:
\begin{eqnarray}
\Omega_\text{LT}&=&\frac{a B^\frac{\gamma -3}{2}}{\rho^5 \sqrt{\Delta}}
\left\{  \cos^2\theta \left[\rho^4 B \left(1-B^{\gamma }\right)+2 a^2 M r \sin^2\theta \right]^2 \right. \nonumber\\
& &+ \left. M^2 \sin^2\theta \Delta  \left(\rho^2-2 r^2\right)^2 \right\}^{\frac{1}{2}},
\label{LT_RJNW}
\end{eqnarray}
with $B = 1 - A/\gamma$. This general relation is applied for deduction of subsequent results.

\begin{figure*}
\mbox{
\subfigure{\includegraphics[width=8.0cm]{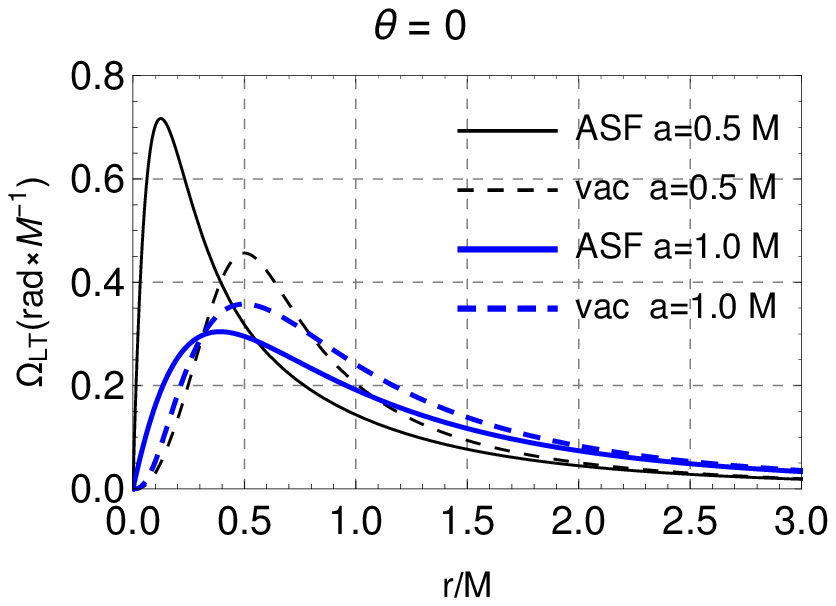}} \qquad \qquad
\subfigure{\includegraphics[width=8.0cm]{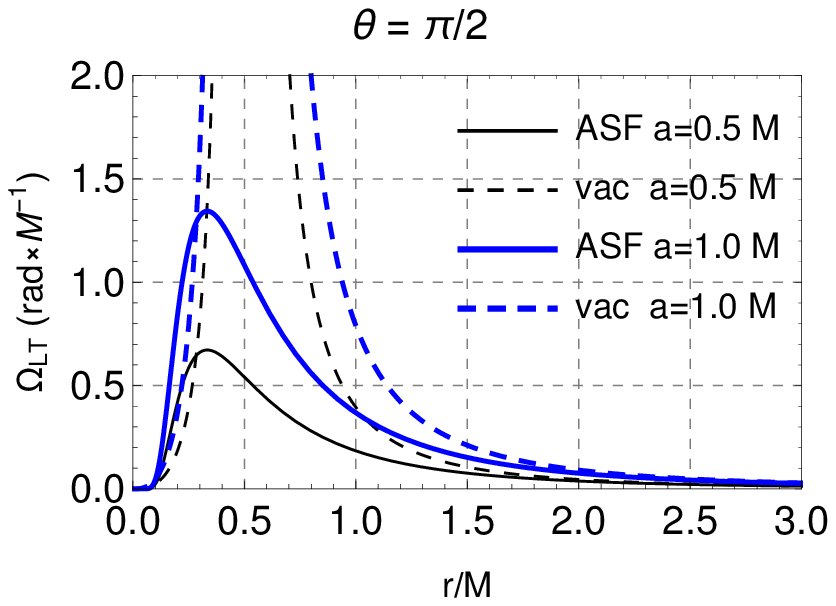}}
}
\caption{Behavior of Lense-Thirring precession frequency (in units of $M^{-1}$) as a function of isotropic radial coordinate $r$ for antiscalar (`ASF') and vacuum (`vac') background, in polar (left, $\theta=0$) and equatorial (right, $\theta=\pi/2$) orbital planes, each for two values of specific angular momentum $a$.}
\label{fig:LT}
\end{figure*}

\subsection{Kerr (vacuum) case } 
\label{sub:kerr_}

In this case one should pose $\gamma =1$ in (\ref{RJNW})-(\ref{defs}), then $\Delta(\gamma =1)=r^2 + a^2 - 2M r$, and so the standard Kerr vacuum metric follows:
\begin{eqnarray}
d{s}^2 &=& \left(1 - A\right) \left(dt-W d\phi \right)^2 \nonumber\\
 &-&\rho^2\left(\frac{dr^2}{\Delta}+d\theta^2+\sin^2\theta d\phi^2\right) \nonumber\\
&+& 2 W(dt-W d\phi)d\phi,
\label{kerr}
\end{eqnarray}
with $\sqrt{-g} = \rho^2 \sin \theta$. In accord with (\ref{magnitudeLT}), the magnitude of the vector (\ref{GenPrec0}) evaluated for the Kerr metric is \cite{0264-9381-31-7-075006}
\begin{eqnarray}
\Omega_\text{LT}(r, \theta) & = & \frac{a M}{\rho^3 \left(\rho^2-2 M r\right)}  \nonumber \\
& \times & \sqrt{4 r^2 \Delta \cos^2\theta +\left(\rho^2-2 r^2\right)^2\sin^2\theta }\,\, ,
\label{LTkerr}
\end{eqnarray}
which also may be obtained from (\ref{LT_RJNW}) by taking $\gamma =1$. As was noted in Section \ref{sec:geodetic}, for correct comparison with antiscalar case one should apply the transformation (\ref{isotropTransform}) to ``isotropic'' coordinates in (\ref{kerr})-(\ref{LTkerr}). Thus, e.g., in (\ref{LTkerr}) one should insert $r\left(1+\frac{M}{2r}\right)^2$ instead of $r$ into $\Omega_\text{LT}(r,\theta)$.

\subsection{Rotating mass in the antiscalar background} 
\label{sec:RAS}
In this case we adopt $\gamma \rightarrow \infty$ in (\ref{RJNW}), and  obtain the rotational generalization of the  antiscalar Papapetrou metric:
\begin{eqnarray}
ds^2 = e^{-A} \left(dt-W d\phi \right)^2 &-&
e^{A} \rho^2\left(\frac{dr^2}{\Delta}+d\theta^2+\sin^2\theta d\phi^2\right)\nonumber\\
&+& 2W(dt-W d\phi)d\phi,
\label{rasf}
\end{eqnarray}
where $A$, $\rho$ and $W$ the same as in (\ref{defs}), but
\begin{equation}
	\Delta = \Delta(\gamma \to \infty) = r^2 + a^2.
	\label{delta}
\end{equation}
The Jacobian of the metric (\ref{rasf}) is
\[
\sqrt{-g}= \rho^2  e^A  \sin\theta.
\]

As a double-check, we have obtained the same result (\ref{rasf}) directly from the Papapetrou metric, applying the Newman-Janis formalism -- see Appendix~(\ref{sec:NJalgorithm}).

As easily seen, with $a \to 0$ (\ref{rasf}) reduces to the Papapetrou solution (\ref{Papa}). Unlike the vacuum Kerr (\ref{kerr}) and scalar rotational JNW (\ref{RJNW}) metric, the solution (\ref{rasf}) is much more simple: it does not contain event horizons and ergospheres, and moreover, for all $r>0$ none of the metric coefficients vanishes or blows up.

The corresponding potential might be obtained from (\ref{KGRJNW}) via the algorithm (\ref{limit}):
\[
\lim_{\gamma\to\infty}\phi = \lim_{\gamma\to\infty}\left[ \frac{\sqrt{1- \gamma^2}}{2}\ln\left(1-\frac{A}{\gamma}\right) \right] = -i\frac{M r}{\rho^2},
\]
i.e., since $\phi \mapsto i\phi$ (cf. (\ref{limphi}) and (\ref{np})), the antiscalar rotational potential proves to be
\begin{equation}
	\phi=\phi(r,\theta) =  \frac{M r}{\rho^2} = \frac{M r}{r^2 +a^2 \cos^2\theta}=\frac{A}{2}.
	\label{potRot}
\end{equation}

In accord with (\ref{magnitudeLT}), the Lense-Thirring vector (\ref{GenPrec0}) evaluated for the metric (\ref{rasf}) has the magnitude:
\begin{eqnarray}
\Omega_\text{LT}&=&\frac{a e^{-A/2}}{ \rho^{5}\sqrt{\Delta}}
\left\{
\cos^2\theta \left[\left(1-e^{-A}\right) \rho^4+2 a^2 M r \sin^2\theta \right]^2 \right. \nonumber\\
 &+& \left. M^2 \sin^2\theta \Delta \left(\rho^2-2 r^2\right)^2 \right\}^{\frac{1}{2}},
\label{LT_RAS}
\end{eqnarray}
which also may be obtained from (\ref{LT_RJNW}) by taking $\gamma \rightarrow \infty$.

Now, this final relation (\ref{LT_RAS}) may be compared with its Kerr-type ``isotropic'' analogue following directly from (\ref{LTkerr}) by applying (\ref{isotropTransform}). Some results of this comparison of vacuum and antiscalar Lense-Thirring effects in strong-field regime (small $r$) are represented in Fig.~\ref{fig:LT}. Manipulation in orientation of orbital plane from $\theta=0$ to $\theta=\pi/2$ shows that singular behavior of $\Omega_\text{LT}$ in vacuum at $r=M/2$ becomes apparent only very close to equatorial plane. Meanwhile, for antiscalar background, $\Omega_\text{LT}$ always behaves monotonically and increases with the growth of specific angular momentum $a$ when moving off the polar plane.

\section{Conclusion}
Working within the standard general relativity algorithm, we have obtained a number of exact results where effects are induced by antiscalar field which (at least in static case) proves to be dynamically and thermodynamically stable. We juxtapose corresponding solutions and their possible observational signatures, both in static and rotational regimes. In particular, we have obtained the new solution (\ref{rasf}) representing the rotational generalization of the spherically symmetric Papapetrou spacetime, as antiscalar counterpart of the vacuum Kerr metric (\ref{kerr}). 

Remarkably, all new analytical results demonstrate, as a rule, practically negligible expected observational differences between the two physically distinct situations -- for objects in vacuum and for those embedded into antiscalar background, when considered in weak-field regime.

Nevertheless, the differences might be reliably traced in strong-field regime, e.g., by shadow imaging of the central object in the Milky Way, as undertaken by the Event Horizon Telescope \footnote{\url{https://eventhorizontelescope.org/}}. Our result is that in static case for a fixed mass of compact object the shadow size is about 5\% larger in antiscalar approach than in vacuum case. We have also confirmed, using isotropic coordinates \emph{ab initio}, the 6.92\% difference in $\omega_\text{isco}$ in vacuum and antiscalar case, predicted earlier by Watt \& Misner \cite{watt1999relativistic}.

When transferring from scalar to antiscalar metrics (static and rotational), the fundamental conclusion is that exactly masses serve as scalar field sources. In the end, the obtained antiscalar solutions are much simpler than their scalar counterparts as they have one free parameter less. At the same time, antiscalar solutions are also simpler than vacuum analogs, as they are deprived of event horizons and ergospheres due to the presence of antiscalar background.

\begin{acknowledgments}
The authors appreciate helpful discussions with Vladimir L. Saveliev. We also thank Adam Crowl for pointing out the interesting reference \cite{boonserm2018exponential}, which echoes some of our results and conclusions (the works are independent as follows from date records). The work is partially supported by the program BR05236322 of the Ministry of Education and Science of the Republic of Kazakhstan.
\end{acknowledgments}

\appendix
\section{On the stability of antiscalar dynamical equation}
\label{sec:Stability}
The dynamical equation for both scalar and antiscalar field is the same Klein-Gordon equation (here, massless):
\begin{eqnarray}
\frac{1}{\sqrt{-g}} \partial_\mu \left( \sqrt{-g} g^{\mu \nu} \partial_\nu \phi \right) =0.
\label{KGap}
\end{eqnarray}
The difference is that in the former case the JNW metric is used, while in the latter case it is the Papapetrou metric. The stability of (\ref{KGap}) in terms of the JNW metric was studied in \cite{sadhu2013naked} using a method where perturbation was introduced via a small positive mass-term. We employ a similar algorithm to the perturbed equation (\ref{KGap}) for antiscalar background with necessarily negative mass-term, as follows from the following consideration.

As was noted in the Introduction, we envisage minimal (anti)scalar field as a limiting case of some massive field. In one of our previous works \cite{Mychelkin2015} we have considered static limit of the usual Einstein-Maxwell equations  $G_{\mu \nu}=\varkappa {T}_{\mu\nu}^\text{EM}$ (EM standing for electromagnetic), with the resulting space EMT-components ${T}_{ij}^\text{EM}(\phi)$ which have exactly negative sign (see \cite{zhuravlev1999latent,marsh2008charge}):
\begin{equation}
	 {G}_{ij}= - \frac{\varkappa}{4 \pi}\left( \phi_{,i} \phi_{,j} -\frac{1}{2} g_{ij} \phi_{,k} \phi^{,k}\right)=-{T}_{ij}^\text{EM}(\phi).
\label{EMTem}
\end{equation}
Here the electrostatic field $\phi$ might be understood, e.g., as the Coulomb field of a positive ($\phi^+$) or negative ($\phi^-$) electric charge. Assuming that there might exist effectively neutral scalar field as a superposition of quasi-static electric fields of the type $\phi \approx \phi^+ +\phi^-$ (e.g., generated by all charged fermions in the Universe), it is admissible to prolongate (\ref{EMTem}) into ordinary 4-dimensional antiscalar EMT (\ref{EMT}) satisfying (\ref{EE3}). Then, it has been found that realistic cosmological solution (reducing at linear approximation to Newtonian gauge) may be obtained for antiscalar field with vanishingly small and negative mass-term fixed by the value of the cosmological constant through the integrability condition $m^2=-\frac{3}{2}\Lambda$ (see \cite{Mychelkin2015}).

Then, considering the minimal antiscalar background related to such cosmological field we apply the algorithm described in \cite{sadhu2013naked} to the perturbed Klein-Gordon equation, but with  negative mass-term:
\begin{eqnarray}
\frac{1}{\sqrt{-g}} \partial_\mu \left( \sqrt{-g} g^{\mu \nu} \partial_\nu \phi \right) =-m^2,
\label{KGmassive}
\end{eqnarray}
which should be analyzed within the antiscalar Papapetrou metric. Following \cite{sadhu2013naked}, we choose the ansatz 
$$
	\phi = \frac{\psi(r)}{r}Y_{lm}(\theta, \phi) e^{i \omega t},
$$
with $Y_{lm}$ the spherical harmonics and tortoise-like coordinate $r_* = r + \frac{2M}{\gamma}\ln\left(\frac{\gamma r}{2M}-1\right)$, which in our case ($\gamma \to \infty$) reduces simply to $r$. In \cite{sadhu2013naked} it was shown that for the JNW-case the spectrum $\omega$ should be real, i.e. $\omega^2 \ge 0$ (see also in \cite{gibbons2005stability}). Because the Papapetrou metric is a particular (limiting) case of the JNW spacetime, this condition holds here as well. Then, the Klein-Gordon equation (\ref{KGmassive}) in antiscalar metric (\ref{Papa}) becomes
$$
	-\frac{d^2 \psi}{dr^2}+\left[ \frac{l(l+1)}{r^2} -m^2 e^{2M/r} \right] \psi = \omega^2 e^{4M/r}\psi.
$$
For the endpoint $r \to \infty$ this reduces to
$$
	-\frac{d\psi^2}{dr^2} -m^2 \psi = \omega^2 \psi,
$$
with the general solution
$$
	\psi = C_1 \cos{kr} + C_2 \sin{kr},
$$
where $k^2=\omega^2+m^2$ is always real. In this case there are no exponentially growing solutions at $r \to \infty$, which implies stability of scalar field under the metric (\ref{Papa}).

\section{Antiscalar thermodynamics}
\label{sec:Thermod}
\subsection{Thermodynamic stability}
The first law of thermodynamics 
\begin{equation}
	dE = dQ-pdV
	\label{1stlaw}
\end{equation}
for quasi-equilibrium states might be recast into the following Gibbs equation (with $s$ being the entropy density and
$q$ the heat flux density):
\begin{equation}
dq = 0 = \Theta d(s/n) = d(\varepsilon/n) + pd(1/n).
\label{Gibbs}
\end{equation}
Following, e.g., Synge \cite{Synge1960} (see \S 14), we express the energy density $\varepsilon = -\partial n / \partial z$ and the pressure $p = n / z$ of perfect fluid as functions of the number density $n = n (z )$ and temperature $\Theta$ through the geometric scalar $z$ related to reciprocal temperature, $z = \Theta^{-1}$. Then (\ref{Gibbs}) transforms into differential equation:
$$
	nn'' + nn'/z -(n')^2 = 0, 
$$
the first integral of which is the barotropic equation of state with constant $w$: 
\begin{equation}
	-w \partial n / \partial z = n/z \qquad \Rightarrow \qquad p=w \varepsilon.
	\label{EoS}
\end{equation}
Integrating once more, we get $n = C \Theta^{1/w}$, and so
\begin{equation}
	 \varepsilon =  \frac{C}{w}\Theta^{1+\frac{1}{w}}, \qquad p = C\Theta^{1+\frac{1}{w}},
	\label{n}
\end{equation}
\begin{equation}
	\frac{s}{k_\texttt{B}} = \frac{dp}{d\Theta} =  C\left(1+\frac{1}{w}\right)\Theta^{1/w} = z(\varepsilon+p),
	\label{s}
\end{equation}	
where, for closed systems, the chemical potential is taken to be zero \footnote{In general, when chemical potential $\mu \ne 0$, the entropy density is $s/k_\texttt{B} = z(\varepsilon+p-\mu n) =  C(1+1/w - \mu\Theta^{-1})\Theta^{1/w}$.} ($k_\texttt{B}$ is the Boltzmann constant). Obviously, for each value of $w$ in (\ref{n})-(\ref{s}) we have, in general, different medium or type of state, and so $C=C(w)$ is the positive integration constant with dimensionality depending on  $w$. Now, from (\ref{Gibbs}) and (\ref{EoS}) it follows that $p = n \partial \varepsilon / \partial n - \varepsilon = w \varepsilon$, i.e. 
\begin{equation}
\frac{\partial \varepsilon}{ \partial n} = (1+w)\frac{\varepsilon}{n}.
	\label{quasieq}
\end{equation}
Thermodynamic stability might be expressed through the known condition (see, e.g., \cite{barboza,kubo1968thermodynamics}):
$$
	\delta^2 E>0 \quad \Rightarrow \quad  \frac{\partial^2 E}{\partial V^2} \ge 0,
$$
i.e. positivity of the second-order derivative of the energy with respect to some extensive parameter (here, volume). Now, using (\ref{1stlaw}) and transforming $$\frac{\partial}{\partial V} \,\, \rightarrow \,\, \frac{\partial}{\partial (1/n)} = -n^2 \frac{\partial}{\partial n}$$ we obtain for $p=w \varepsilon$:
$$
	\frac{\partial^2 E}{\partial V^2} = -\frac{\partial p}{\partial V} > 0 \quad \Rightarrow \quad n^2 \frac{\partial \varepsilon}{\partial n }w>0,
$$
which, by applying (\ref{quasieq}), yields the final condition of stable equilibrium for systems with the state parameter $w$:
\begin{equation}
	w(w+1) n \varepsilon >0.
	\label{condnes}
\end{equation}
According to (\ref{n}), $\varepsilon$ might be positive or negative depending on the sign of $w$. Then, for $\varepsilon > 0$ (\ref{condnes}) implies
\begin{equation}
	w(w+1)>0,
	\label{wgz}
\end{equation}
while for $\varepsilon <0$ we get
\begin{equation}
	w(w+1)<0.
	\label{wlz}
\end{equation}
Comparing the standard scalar field EMT (\ref{EMT}) with the perfect fluid EMT
$$
	{T}_{\mu\nu}^\text{pf} = (\varepsilon + p)u_\mu u_\nu - p {g}_{\mu\nu},
$$
we adopt $u_\mu = \phi_\mu /\sqrt{\phi_\alpha \phi^\alpha}$ for timelike gradient $\phi_\mu$ ($\phi_\alpha \phi^\alpha>0$, $u_\alpha u^\alpha = 1$), and $u_\mu = \phi_\mu /\sqrt{-\phi_\alpha \phi^\alpha}$ for spacelike $\phi_\mu$ ($\phi_\alpha \phi^\alpha<0$, $u_\alpha u^\alpha = -1$). Then, we obtain the equation of state parameter $w = 1$ for timelike and $w=-1/3$ for spacelike gradient. Thus, according to (\ref{wgz}) and (\ref{wlz}), scalar and antiscalar backgrounds for both $w = 1$ and $w=-1/3$ might exist as thermodynamically stable media. 

However, the condition (\ref{condnes}) is necessary but insufficient, since we haven't yet included gravity into equilibrium thermodynamics. In general, such inclusion implies the existence of a time-like Killing field $\xi^\mu = \xi u^\mu$, with standard modulus $\xi = \xi_\mu u^\mu = \sqrt{g_{00}}$, which, on the other hand, is equal (up to  some general relativistic invariant $\Theta_0$) to the reciprocal temperature,  $\xi =\Theta_0 z= \Theta_0/\Theta$, so that $\Theta_0 = \Theta \sqrt{g_{00}}$, thus taking gravitational redshift factor into account \cite{Landau}. With gravity present, we require for all quantities to be general-relativistic invariants, thus replacing  $\Theta$ by $\Theta_0$.

Now, for spacelike case, taking the trace of standard Einstein's equations ${G}_{\mu\nu}~=~\varkappa {T}_{\mu\nu}^\text{pf}$ with (\ref{n}), we get:
\begin{equation}
-R = -\varkappa \left(\varepsilon + 5p\right) = -\varkappa C  \frac{1+5w}{w}\Theta_0^{1+\frac{1}{w}},
\label{traceSL}
\end{equation} 
i.e., since in this case $w=-1/3$,
\begin{equation}
R = \frac{2 \varkappa C}{\Theta_0^2} > 0,
\label{condSL}
\end{equation}
positing the non-negativeness of the square of temperature. 
However, for the scalar JNW solution (\ref{JNW}) the Ricci scalar
\begin{equation}
R = \frac{2 G^2 M^2  \left(\gamma ^2-1\right) }{\gamma^2 c^4 r^4} \left( 1- \frac{2GM}{\gamma c^2 r}  \right)^{\gamma-2}
\label{RicciScJNW}
\end{equation}
is negative for all $0<\gamma<1$, and thus contradicts (\ref{condSL}).

Transfer to antiscalar mode is equivalent to the change of the sign of the trace of ${T}_{\mu\nu}^\text{sc}$, i.e. $\phi_\alpha \phi^\alpha <0 \to \phi_\alpha \phi^\alpha >0$, which, as mentioned above, implies effective state with $w=1$. In this case we have
\begin{equation}
-R = \varkappa \left(\varepsilon -3p\right) = \varkappa C  \frac{1-3w}{w}\Theta^{1+\frac{1}{w}}_0,
\label{traceTL}
\end{equation}
i.e.
\begin{equation}
R = 2 \varkappa C \Theta^2_0 > 0,
\label{condT}
\end{equation}
again positing the non-negativeness of the square of invariant temperature as a sufficient condition of general-relativistic stability. The conclusion is that (\ref{condT}) is satisfied by antiscalar Papapetrou solution (\ref{Papa}) which is self-consistent due to the proper sign of the Ricci scalar:
\begin{equation}
R = 2 \frac{ G^2 M^2}{c^4 r^4} \exp\left(\frac{-2GM}{c^2r}\right) > 0.
\label{condPapa}
\end{equation}
On the contrary, for the scalar JNW solution (\ref{JNW}) the Ricci scalar (\ref{RicciScJNW})
is negative for all $0<\gamma<1$, and thus contradicts (\ref{condT}). So, only antiscalar stationary state background is attainable within thermodynamically consistent general-relativistic approach.

\subsection{Relation to BH-thermodynamics}

The difference between conditions (\ref{condSL}) and (\ref{condT}) is that the first leads to exotic thermodynamics (with negative $w$, as for tachyons and strings) while the second reduces (as a particular case) to the black hole thermodynamics, as shown below.

In accord with (\ref{condT}) and (\ref{condPapa}), the local temperature ($\Theta = \Theta_0/\sqrt{g_{00}}$) of antiscalar background in the Papapetrou metric is:
\begin{equation}
	\Theta(r) =\frac{1}{2\sqrt{2 \pi}} \sqrt{ \frac{G}{C} } \frac{M}{r^2}.
	\label{loctemp}
\end{equation}
On equipotential surface with $r=r_g=2GM/c^2$ we get for the value of local temperature $\Theta$ at this scale:
\begin{equation}
	\Theta(r_g) = \frac{c^4}{8\sqrt{2 \pi}\sqrt{CG^3}} \frac{1}{M},
	\label{BHtemp}
\end{equation}
which is similar to the Hawking black hole temperature:
\begin{equation}
	\Theta_{\text{BH}} = k_\text{B} T_{\text{BH}} = \frac{\hbar c^3}{8\pi G}\frac{1}{M}.
	\label{tbh}
\end{equation}
Comparison of (\ref{BHtemp}) with (\ref{tbh}) yields corresponding value for $C$:
\begin{equation}
	C = C(w=1) = \frac{\pi c^2}{2\hbar^2 G}.
	\label{Cw1}
\end{equation}
In a sense, this situation resembles that of the Stefan-Boltzmann law for radiant emittance $ j_\text{rad} = \sigma T^4$ which had been found classically, and then the phenomenological constant $\sigma$ was estimated by quantum methods.

So, the corresponding local densities (\ref{n})-(\ref{s}) are well-defined for $w=1$ on each equipotential surface with $C$ and $\Theta$ given by (\ref{Cw1}) and (\ref{loctemp}):	$n = C \Theta,  \varepsilon = C \Theta^2, p = C \Theta^2, s = k_\texttt{B} ( 2 C \Theta).$
In particular, full entropy $S(r_g)$, related to the domain inside $r=r_g$ is
\begin{eqnarray}
	S(r_g) &=& \int{s_\mu dV^\mu} = 4 \pi \int_0^{r_g}{s(r)r^2dr} \nonumber\\
	 &=&8 \pi C k_\texttt{B} \int_0^{r_g}{\Theta(r)r^2 dr} = k_\texttt{B}\frac{4 \pi  G }{\hbar c} M^2,\nonumber
\end{eqnarray}
where $s_\mu = su_\mu$ and $dV^\mu = u^\mu d^3V$. The last result coincides with the well-known relation for black hole entropy $S_\text{BH} = k_\texttt{B} A/(4\ell_P^2)= k_\texttt{B} (\pi r_g^2/\ell_P^2)$, with $A$ being the area of the horizon and $\ell_P = \sqrt{ \hbar G/c^3}$ the Planck length. Thus, the antiscalar thermodynamics includes traditional black hole thermodynamics (at $r=r_g$) as a particular case (cf. \cite{cadoni2015asymptotically}). This might serve as another argument in favor of the physical relevance of antiscalar background.

\section{Derivation of the antiscalar rotational metric using the Newman-Janis algorithm}
\label{sec:NJalgorithm}

Following the procedure described in \cite{Agnese1985}, we begin with the static antiscalar Papapetrou metric (\ref{Papa}) which in the radiation form can be written as
$$
	ds^2 = e^{\alpha(r)} du^{2} + 2du dr - e^{-\alpha(r)} r^2 \left( d \theta^2 + \sin^2 \theta d \phi^2  \right),
$$
where the new time coordinate $u$ is defined as
$$
	du = dt - e^{-\alpha(r)}dr, \qquad e^{\alpha(r)} = e^{-2M/r}.
$$
Now, in accord with the Newman-Janis algorithm \cite{NewmanJanis65}, we write this metric in terms of complex null tetrad:
\begin{eqnarray*}
	g^{\mu \nu} &=& l^\mu n^\nu + l^\nu n^\mu - m^\mu \bar{m}^\nu - m^\nu \bar{m}^\mu, \\
	l^\mu &=& (0,1,0,0), \qquad n^\mu = \left( 1, -e^{\alpha(r)}/2, 0, 0 \right),\\
	m^\mu &=& \frac{1}{\sqrt{2} r e^{-\alpha(r)/2}} \left( 0, 0, 1, \frac{i}{\sin \theta}  \right),\\
	\bar{m}^\mu &=& \frac{1}{\sqrt{2} r e^{-\alpha(r)/2}} \left( 0, 0, 1, -\frac{i}{\sin \theta}  \right),
\end{eqnarray*}
where $l$ and $n$ correspond to principal null vectors of the Weyl tensor in coordinates $x^\mu = (u,r,\theta, \phi)$.

To transfer to real coordinates, the following complex transformation is applied:
\begin{eqnarray*}
	u' &=& u- ia\cos\theta,\\
	r' &=& r + ia\cos\theta,\\
	\theta' &=& \theta,\\
	\phi' &=& \phi, 
	\label{zzz4}
\end{eqnarray*}
which leads to new null tetrad (we drop the primes):
\begin{eqnarray*}
l^\mu &=& (0,1,0,0),  \qquad n^\mu = \left( 1, -e^{\alpha(r,\theta)}/2, 0, 0 \right), \\
m^\mu &=& \frac{1}{\sqrt{2}(r+ia\cos\theta)  e^{-\alpha(r,\theta)/2}} \left( ia\sin\theta, -ia\sin\theta, 1, \frac{i}{\sin \theta}  \right), \\
\bar{m}^\mu &=& \frac{1}{\sqrt{2}(r-ia\cos\theta)  e^{-\alpha(r,\theta)/2}} \left( -ia\sin\theta, ia\sin\theta, 1, \frac{-i}{\sin \theta}  \right), 
\label{zzz5}
\end{eqnarray*}
where now we can write, using (\ref{potRot}),
$$
	e^{\alpha(r,\theta)} = e^{-\frac{2Mr}{\rho^2}} = e^{-2 \phi(r,\theta)}, \quad  \rho^2 = r^2+a^2 \cos^2 \theta
$$
(the potential $\phi=\phi(r,\theta)$ not to be confused with the coordinate $\phi$).
Performing the coordinate transformation
\begin{eqnarray*}
	du &=& d\hat{t} - \left(  \frac{e^{-\alpha(r,\theta)} \rho^2 + a^2 \sin^2 \theta }{r^2 + a^2} \right)dr,\\
	d \phi &=& d\hat{\phi} - \left( 	\frac{a}{r^2 + a^2}  \right) dr
	\label{zzz7}
\end{eqnarray*}
we obtain, dropping the hats, the following line element:
\begin{eqnarray*}
	ds^2 &=& e^{-2 \phi} dt^2 - \frac{dr^2}{e^{-2 \phi} \left( 1 + \frac{a^2\sin^2\theta}{\rho^2} \right)} \\
	&+& 2\left( 1-e^{2 \phi}  \right) a \sin^2\theta dt d \phi \\
	&-& e^{2 \phi} \rho^2 \left\{ d\theta^2 + \left[ 1 + \frac{(2-e^{-2 \phi})a^2 \sin^2\theta}{e^{2 \phi} \rho^2} \right] \sin^2 \theta d \phi^2 \right\}.
\end{eqnarray*}
Then in final form the sought-after  Kerr metric analog for anti\-scalar background might be rewritten as:
\begin{eqnarray}
ds^2 &=& e^{-2 \phi(r,\theta)} \left(dt- a\sin^2\theta d\phi \right)^2 \nonumber\\
&-& e^{2 \phi(r,\theta)} \rho^2\left(\frac{dr^2}{\Delta}+d\theta^2+\sin^2\theta d\phi^2\right)\\
&+& 2a\sin^2 \theta(dt- a \sin^2\theta d\phi)d\phi, \nonumber
\end{eqnarray}
with $\Delta=r^2 + a^2$. This expression coincides with (\ref{rasf}).

\bibliography{references}

\end{document}